\documentclass{JAIS}

\journal{JAIS-ID}
\vol{2022}

\received{xx January 2018}
\published{xx March 2018}

\def\be{\begin{equation}}
\def\ee{\end{equation}}
\def\bea{\begin{eqnarray}}
\def\eea{\end{eqnarray}}

\usepackage{caption}
\usepackage{lineno}
\usepackage{times}
\usepackage{hyperref} 

\begin{document}

\title{Simulation Studies of a Novel Muography Detector for the Great Pyramids}

\author{
S. Aly\auno{3,5},
Y. Assran\auno{4,6}, 
A. Bonneville\auno{8,10}, 
B. ElMahdy\auno{3,4}, 
R.T. Kouzes\auno{10},
A. Lintereur\auno{9},
A. Mahrous\auno{2},
I. Mostafanezhad\auno{7},
R. Pang\auno{7},
B. Rotter\auno{7},
F. Snigdha\auno{7},
M. Tytgat\auno{1},
M. N.Yaseen\auno{3}
}
\address{$^1$Ghent University, Dept. of Physics and Astronomy, Ghent, Belgium}
\address{$^2$Institute of Basic and Applied Science, Egypt-Japan University of Science and Technology, Alexandria, Egypt}
\address{$^3$Department of physics, Faculty of Science, Helwan University, Cairo, Egypt}
\address{$^4$Centre for Theoretical Physics, The British University in Egypt, El Sherouk City, Cairo, Egypt} 
\address{$^5$Canadian International college, Sheikh Zayed Campus, Giza, Egypt}
\address{$^6$Suez university, El Salam City, Suez-Cairo Road, Suez, Egypt}
\address{$^7$Nalu Scientific, Honolulu, Hawaii, USA}
\address{$^8$Oregon State University, Corvallis, Oregon , USA}
\address{$^9$Penn State University, State College, Pennsylvania, USA}
\address{$^{10}$Pacific Northwest National Laboratory, Richland, Washington, USA}

\begin{abstract}
Muography is an imaging technique that can be used to examine the interior structure of large-size objects. The technique is based on measurements of the absorption of cosmic-ray muons passing through the object under study. 
 
Muographic imaging was already successfully applied to the discovery of a new void inside the Great Pyramid of Khufu at Giza, Egypt. 
With the aim of studying the Pyramid of Khafre, the second largest pyramid at Giza, a new muography detector is currently being constructed. In this paper we report on the development of a corresponding new simulation framework of the Great Pyramids and the detector setup.  
This simulation will serve as a basis to develop the data and image reconstruction algorithms to be used during our future muography campaign in the Pyramid of Khafre, and will allow us to study the relevant experimental conditions at the site. 
\end{abstract}

\maketitle



\section{Introduction}
Cosmic rays are high-energy radiation, mainly originating from outside our solar system and from distant galaxies. When they strike our Earth's atmosphere, their interactions with air result in showers of secondary particles that travel and interact further in the atmosphere. Among these secondary particles, the abundantly produced charged pions and kaons will primarily decay into muons that, given their long lifetime and relatively low interaction cross section, can penetrate a large amount of material and still reach ground level. This feature proves to be useful, allowing cosmic muons to be used as imaging probes to study the inner structure of large objects up to the km-scale, for which other traditional imaging techniques cannot be used. 
This method, generally called muon radiography, or muography for short, is a non-destructive technique based on absorption (or transmission) measurements of cosmic-ray muons which can be used for a variety of imaging applications~\cite{BONOMI2020103768}.  

One such applications of muography is the study of archaeological heritage, with the particular example of the imaging of the interior structure of  pyramids. The precise internal layout of the Egyptian pyramids still remains a complex mystery. Pyramids may contain several voids linked to different corridors with the aim of protecting the king's mummy during his life after death. 
In muographic imaging of a pyramid's inner layout, the downward going flux of cosmic muons passing through the pyramid can be detected by detectors placed near or inside the structure. Compared to free sky measurements, the muon flux transmitted through the pyramid will reflect the average density of the material along the muon's trajectory, and may hence reveal voids or cracks in the structure that have not been previously found. This technique of using muon detectors to search for unknown voids or other features inside the pyramids has the advantage of not causing any damage and allows probing the entire pyramid structure with a single setup. To derive a true 3D tomographic image of a pyramid one does need to perform measurements from multiple viewpoints, i.e., by either moving the detector to different positions or by operating multiple detectors simultaneously at different locations around or inside the structure. 

Muographic imaging of pyramids was initially performed in the late 1960s in the Pyramid of Khafre at Giza, where no hidden chambers were discovered~\cite{alvarez1970}. Later on, muographic measurements were also performed on the Teotihuacan Pyramid of the Sun in Mexico, but again, no new structures were discovered~\cite{Aguilar2013}. Nonetheless, the more recent ScanPyramids mission led to the discovery of a hidden void in the Pyramid of Khufu using a combination of different detector setups and technologies~\cite{Morishima2017}, which constituted a breakthrough moment for the use of muographic techniques in archaeological studies. 

Following the success of the ScanPyramid mission, this project now aims to perform new muographic measurements in the second largest pyramid at Giza, the Pyramid of Khafre or of Chephren. 
A new scintillator-based muon detector is currently under construction~\cite{Kouzes2022}. In this paper we report on the development of a corresponding new simulation of the Great Pyramids and the detector setup. 

This simulation framework will serve as a basis to develop the data reconstruction and imaging algorithms to be used during our future muography campaign at the Pyramid of Khafre. It will also allow us, e.g., to study different options to position the detector and to assess the background conditions and expected muon rates.
The simulation of cosmic-ray muons is performed with the CRY generator~\cite{CRY}

package. 
The GEANT4 full simulation package~\cite{Agostinelli2003} is used to model the pyramid and the detector, to handle the muon transportation and to study the cosmic muon interactions in the pyramid material and the detector setup. 

\section{Simulation of the Pyramid}

The pyramids at Giza are made of limestone where muons have a stopping power of 1.686~MeV.g/cm$^2$ and a critical energy of 630~GeV~\cite{Groom2001}.
The actual height of the Pyramid of Khafre~\cite{pyramidlehner} is 136.4~m, with a base length of 215.3~m, and  
is built out of limestone blocks weighing over 2 tons each. The pyramid slope rises at a 53° 10' angle, which is somewhat steeper than its taller neighbor, the Pyramid of Khufu that has an angle of 51°50'40". 
Internally, the pyramid has two entrances leading to the burial chamber, i.e., one is at ground level, while another is opening 11.54~m 
up the face of the pyramid. These corridors are not aligned with the center line of the pyramid, but are offset to the east by 12~m. 
The lower descending passageway is first descending, then going horizontal before ascending again to join the horizontal corridor that leads to the burial chamber. The main burial chamber with a rectangular ground surface of about $14.2\times 5$~m$^2$ and a height of about 6.8~m was carved out of a pit in the bedrock. 
\begin{figure}[htbp]
\centering
 \includegraphics[width=.45\textwidth, origin=c]{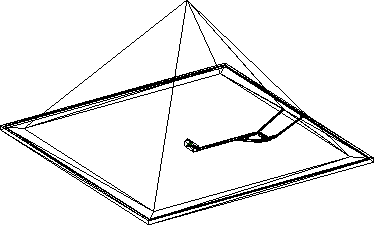}\qquad
 \includegraphics[width=.45\textwidth, origin=c]{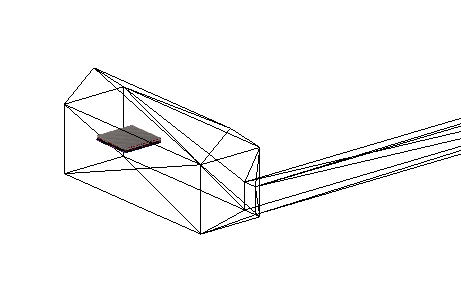}
\caption{\label{fig:py} (left) The model of the Pyramid of Khafre imported into GEANT4. (right) GEANT4 simulation of the muon detector setup inside the main burial chamber of the pyramid.}
\end{figure}

\subsection{Importing the Pyramid Geometry into GEANT4} 

An STL (Standard Tessellation Language) model of the pyramid~\cite{grummufish} was converted into an ASCII format, which was then using the CADMesh tool~\cite{Poole2012} imported into GEANT4 as a solid. The GEANT4 material assigned to the pyramid was Limestone CaCO$_3$, as defined by its bulk density of 2.75~g/cm$^3$ and element mass fractions of $\mbox{Ca}=0.4$, $\mbox{C}=0.12$ and $\mbox{O}=0.48$.

The interior layout of this pyramid as currently included in the simulation is shown in Figure~\ref{fig:py} (left). The model includes the burial chamber and the two entrances that lead to it.

\section{Simulation of the Muon Detector}

The new pyramid muon detector, described in more detail in~\cite{Kouzes2022}, is made of two solid plates of EJ-200 scintillator~\cite{eljentechnology2021}, each of which have orthogonal grooves on either side of the plate containing BCF-92 wavelength shifting fibers (WLS)~\cite{crystals2021}. The simulation of each plate used two plates of half the total thickness. The two solid plates of scintillator will be seperated by some distence to form two x-y layers that can reconstruct each muon trajectory. The fibers embedded in the plate are individually connected to Hamamatsu S14160 Multi-Pixel Photon Counters (MPPC)~\cite{hamamatsu2021} for the light readout. Figure 2 shows on the left the concept for the muon detector where two planes each determine an x and y coordinate that can be used to determine the muon trajectory. On the right of Figure 2 is a photograph of one scintillator plate with the fibers on each face before it was coated with a non-reflective, black paint covering.

The detector layout was drawn using AutoCAD2018~\cite{AutoCad} and its different parts, i.e. scintillator plates and fibers, were separately extracted to binary STL files, which were then each converted to an ASCII STL form to be fed into GEANT4 through the CADMesh~\cite{Poole2012} utility as a solid structure. In GEANT4 the parts were then linked to their corresponding specified material. Figure 3 shows the AutoCad structure of one of the solid scintillator plates with the orthogonal grooves on the opposite faces where the fibers were inserted. The figure shows how the simulation used two parts to make up each plate. Figure 4 shows the Geant4 simulation of the grid of fibers embedded in the scintillator plate.

The EJ-200 scintillator from Eljen Technology has a long optical attenuation length and fast timing, and emits photons in the range of approximately $400-500$~nm, with a maximum emission wavelength at 425~nm. The detailed EJ-200 material specification was included in the GEANT4 material description.
The scintillator plate dimensions are $2\times 61\times 61$~cm$^3$, with 2~mm wide and 3~mm deep grooves spaced 1~cm apart.

\begin{figure}
\centering
    \includegraphics[width=.395\textwidth]{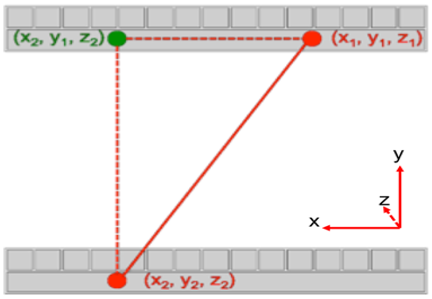}\qquad
    \includegraphics[width=.505\textwidth]{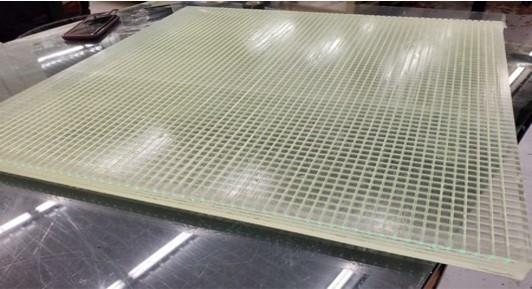}
\caption{\label{fig:BMD} (left) Muon detector concept. (right) Scintillator plate with fibers inserted in the grooves~\cite{Kouzes2022}.}
\end{figure}

The BCF-92 wavelength shifting fibers from Saint-Gobain Crystals absorb photons in the range of 359-458~nm and re-emit in the range of 465-502~nm. These fibers are aligned in the scintillator plate grooves as 60 parallel fibers. They transfer the scintillation signal to the MPPC to generate the final electrical signal related to the incident radiation. In GEANT4 the NIST material G4\_PLEXIGLASS was used for the fibers, and the material was further characterized with its optical spectrum, refractive index and decay time as specified in the manufacturer's product sheet~\cite{crystals2021}. 
 
\begin{figure}[htbp]
\centering 
 \includegraphics[width=.45\textwidth]{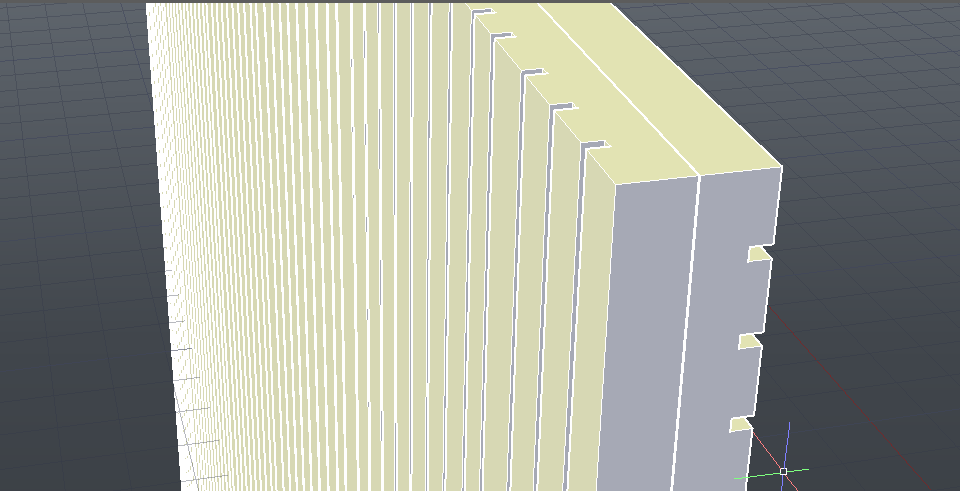}
\caption{\label{fig:EJ} AutoCad structure for the simulated EJ-200 scintillation plate.}
\end{figure}

 \begin{figure}[htbp]
\centering 
 \includegraphics[width=.45\textwidth]{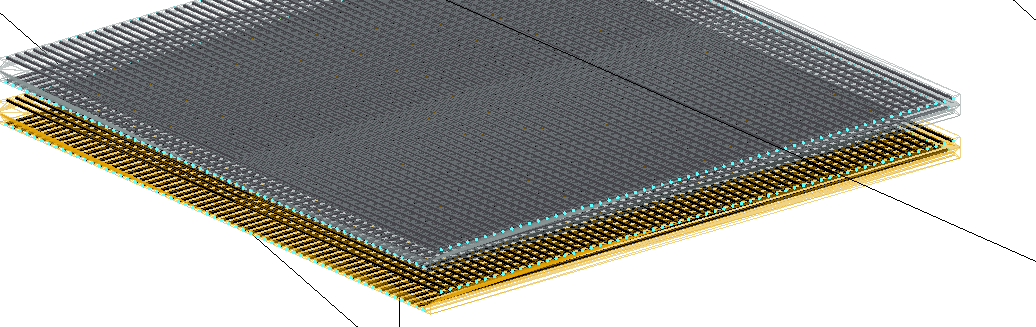}
 \qquad
\includegraphics[width=.45\textwidth,origin=c]{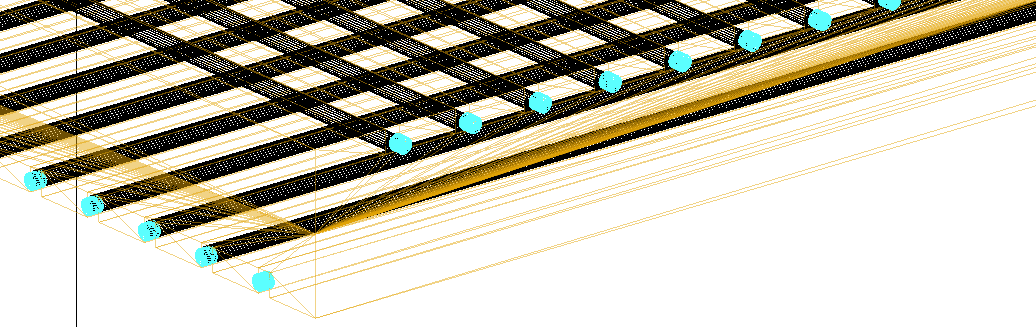}
\caption{\label{fig:grid} GEANT4 simulation of the WLS fibers grid.}
\end{figure}





\begin{figure}[htbp]
\centering
\includegraphics[width=.45\textwidth]{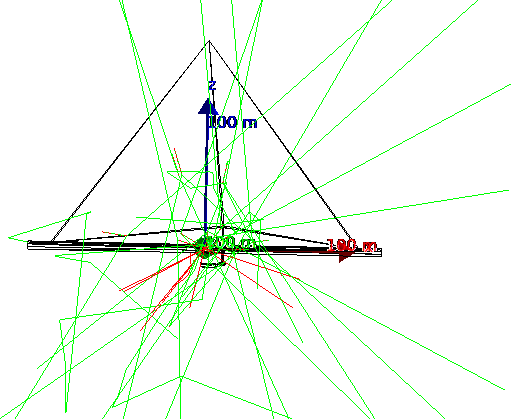}
\caption{\label{fig:KhafreCRY1}Simulation with the CRY generator of cosmic-ray muons hitting the Pyramid of Khafre.}
\end{figure}

A possible location for the detector setup during the muographic campaign of the pyramid could be the main burial chamber as shown in Figure~\ref{fig:py} (right), although also other possible locations will be investigated. With the basic elements of the GEANT4 simulation in place, initial studies of the cosmic-ray muon interactions with the pyramid structure and the detector are currently ongoing, as shown in Figure~\ref{fig:KhafreCRY1}. Next step will include the digitization of the simulated detector response, and the development of the muon track reconstruction algorithm. 










\section{Summary}
A GEANT4-based simulation framework is being developed for a new muographic imaging campaign targeting the Pyramid of Khafre at Giza in Egypt. An STL model of the pyramid including the known internal passageways and burial chamber was imported into GEANT4. 
The detector setup which is currently under construction has a scintilator-based planar layout with a MPPC readout. The  geometry of the muon detector is modeled in GEANT4 and includes the detailed layout of the scintillator plates, wavelength shifting fibers and MPPCs. The CRY package is used for the generation of the cosmic muons, while the muon transport through the pyramid and the interactions with the detector is handled by GEANT4.  

With the basic parts of the simulation in place, the muon track reconstruction algorithms will now be developed that can be applied to both simulated, and later on, measured data. Once the detector itself has been assembled, initial measurements in the lab with cosmic muons will allow a more detailed comparison between measured and simulated data. This will enable the implementation and tuning of the digitization of the detector signal response in the simulation.

\section*{Acknowledgements}
The detector is built through the fund provided by the Egyptian Academy of Scientific Research and Technology (ASRT), Project ID: 6379. PNNL is operated for the US Department of Energy by Battelle under contract DE-AC05-76RLO 1830. PNNL-SA-167973. 

No conflict of interest exists.

\bibliographystyle{unsrt}
\bibliography{posterBibFile.bib}

\begin{thebibliography}{10}

\bibitem{BONOMI2020103768}
G.~Bonomi, P.~Checchia, M.~D’Errico, D.~Pagano, and G.~Saracino.
\newblock Applications of cosmic-ray muons.
\newblock {\em Progress in Particle and Nuclear Physics}, 112:103768, 2020.

\bibitem{alvarez1970}
L.W. Alvarez et~al.
\newblock Search for hidden chambers in the pyramids: The structure of the
  second pyramid of giza is determined by cosmic-ray absorption.
\newblock {\em Science}, 167:832--839, 1970.

\bibitem{Aguilar2013}
S.~Aguilar et~al.
\newblock Search for cavities in the teotihuacan pyramid of the sun using
  cosmic muons: preliminary results.
\newblock In {\em Proceedings of the 33rd International Cosmic Ray Conference
  (ICRC2013), Rio De Janeiro, Brazil}, 2013.

\bibitem{Morishima2017}
K.~Morishima et~al.
\newblock {Discovery of a big void in Khufu's Pyramid by observation of
  cosmic-ray muons}.
\newblock {\em Nature}, 552(7685):386--390, dec 2017.

\bibitem{Kouzes2022}
R.~Kouzes et~al.
\newblock Novel muon tomography detector for the pyramids.
\newblock {\em submitted to JAIS}, (2022).

\bibitem{CRY}
C.~Hagmann, D.~Lange, and D.~Wright.
\newblock {Cosmic-ray shower generator (CRY) for Monte Carlo transport codes}.
\newblock In {\em 2007 IEEE Nuclear Science Symposium Conference Record},
  volume~2, pages 1143--1146. IEEE, 2007.

\bibitem{Agostinelli2003}
S.~Agostinelli, J.~Allison, et~al.
\newblock {GEANT4 - A simulation toolkit}.
\newblock {\em Nuclear Instruments and Methods in Physics Research, Section A:
  Accelerators, Spectrometers, Detectors and Associated Equipment}, 506(3),
  2003.

\bibitem{Groom2001}
D.E. Groom, N.V. Mokhov, and S.I. Striganov.
\newblock {Muon Stopping Power and Range Tables 10 MeV–100 TeV}.
\newblock {\em Atomic Data and Nuclear Data Tables}, 78(2):183--356, jul 2001.

\bibitem{pyramidlehner}
M.~Lehner.
\newblock {\em The Complete Pyramids: Solving the Ancient Mysteries}.
\newblock Thames \& Hudson, 1st ed. edition, 1997.

\bibitem{grummufish}
{Wikimedia Commons - File:Pyramid of Khafre.stl}.
\newblock
  \url{https://commons.wikimedia.org/wiki/File:Pyramid\_of\_Khafre.stl}.

\bibitem{Poole2012}
C.~M. Poole, I.~Cornelius, J.~V. Trapp, and C.~M. Langton.
\newblock {A CAD interface for GEANT4}.
\newblock {\em Australasian Physical and Engineering Sciences in Medicine},
  35(3):329--334, sep 2012.

\bibitem{eljentechnology2021}
{EJ-200, EJ-204, EJ-208, EJ-212 - Plastic Scintillators - Eljen Technology}.
\newblock
  \url{https://eljentechnology.com/products/plastic-scintillators/ej-200-ej-204-ej-208-ej-212}.

\bibitem{crystals2021}
{Saint-Gobain Crystals - Scintillation Materials - Fibers}.
\newblock
  \url{https://www.crystals.saint-gobain.com/products/scintillating-fiber}.

\bibitem{hamamatsu2021}
{Hamamatsu S14160/S14161 series Multi-Pixel Photon Counter (MPPC) Product
  Sheet}.
\newblock
  \url{https://www.hamamatsu.com/content/dam/hamamatsu-photonics/sites/documents/99\_SALES\_LIBRARY/ssd/s14160\_s14161\_series\_kapd1064e.pdf}.

\bibitem{AutoCad}
{AutoCAD Web App - Online CAD Editor \& Viewer | Autodesk.}
\newblock \url{https://web.autocad.com/acad/me/drawings/714839992/editor}.

\end{thebibliography}

\end{document}